\documentclass[12pt,a4paper]{article}

\usepackage{amsmath,amsfonts,amssymb,bm}

\usepackage{graphicx} 

\usepackage[hang,normalsize,bf]{subfigure} 

\begin{document}

\begin{center}

{\Huge \bf Fermi motion and \\[5pt] nuclear modification factor}

\vspace {0.6cm}

{\large A. Szczurek $^{1,2}$ and A. Budzanowski $^{1}$}

\vspace {0.2cm}

$^{1}$ {\em Institute of Nuclear Physics\\
PL-31-342 Cracow, Poland\\}
$^{2}$ {\em University of Rzesz\'ow\\
PL-35-959 Rzesz\'ow, Poland\\}

\end{center}

\begin{abstract}
It has been argued recently that the so-called nuclear modification
factor ($R_{AA}$) is an observable useful for identifying
the quark-gluon plasma.
We discuss the effect of Fermi motion in nuclei
on $R_{AA}$ at CERN SPS and BNL RHIC energies.
Contrary to the simple intuition, rather large
effects are found for CERN SPS.
The Fermi motion in nuclei contributes significantly to the Cronin
effect.
The effect found is qualitatively similar to the one observed
experimentally at CERN energies and similar to the one
obtained in the models of multiple scattering of initial partons.
We predict different size of the effect for different types of hadrons,
especially at low energies.
\end{abstract}

\section{Introduction}

It was realized only recently that the nuclear modification factor defined as
\begin{equation}
R_{AA}^h(y,p_t;W) = \frac{\frac{d \sigma^{AA \to h}}{d y d^2 p_t}(y,p_t;W)}
{N_{bin} \frac{d \sigma^{pp \to h}}{d y d^2 p_t}(y,p_t;W)} \;
\label{R_AA}
\end{equation}
may be a very useful quantity which could shed more light on the dynamics
of high-energy nuclear collisions
\cite{Wang2000,WW2001,GLV02,VG02,ZFPBL2002,BKW03}.
In the above definition
$y$, $p_t$ and $W$ are hadron rapidity, transverse momentum, and
nominal $NN$ center-of-mass energy, respectively.
$N_{bin}$ is the number of binary
nucleon-nucleon collisions usually calculated in the Glauber formalism.
This brings in a small model dependence in the above definition.
The index $h$ indicates the measured hadron.

Quite different values of nuclear modification factors have been
observed at CERN SPS \cite{WA98_pi0} and at the Brookhaven RHIC
\cite{STAR_RAA,PHENIX_RAA,PHOBOS_RAA,BRAHMS_RAA}.
It is speculated that the inequality
\begin{equation}
R_{AA}^h(RHIC) < R_{AA}^h(SPS)
\end{equation}
is due to the creation at RHIC of a dense nuclear medium, a new state
of matter, perhaps the
long seeked quark-gluon plasma
\cite{VG02,dEnterria,Reygers}.
Various nuclear effects like shadowing,
parton multiple scattering or final state quenching
\cite{WW2001,GLV02,VG02} modify
the dependence of $R_{AA}$ on the transverse momentum
and collision energy in quite different ways.
In Ref.\cite{WW2001} the data from CERN SPS were
interpreted within the picture of Glauber multiple
scattering of initial partons (see \cite{parton_rescattering}).
It is not easy to make a priori predictions based on this effect
as one has to model parton-nucleon interaction. The proposed schematic
models are flexible enough so that a rough description
of the data is possible.

In a partonic approach improved for including internal transverse
momenta \cite{Wang2000,ZFPBL2002}, the average transverse momentum
of partons in nuclear collisions is increased compared to the one
in the elementary collisions
\begin{equation}
\langle p_t^2 \rangle_{pA} = \langle p_t^2 \rangle_{pp}
 + \Delta p_t^2(A,b,...) \; .
\end{equation}
The correction term used in the phenomenological analyses
is typically a fraction of GeV, i.e. of the order of
the nuclear Fermi momentum.

At intermediate energies the inclusion of
momentum distributions in nuclei quantifying the nucleon motion
is crucial for the production of kaons \cite{CBMNSW90,CDJKRSS94,Sibirtsev95},
eta mesons \cite{CBVW91} and antiprotons \cite{SP-MG90}
in p+A collisions.
It is well known that at
subthreshold and near-to-threshold collisions the Fermi motion
leads to a huge enhancement of hadron production cross sections
based on the first collision model.
At the lower energies the inclusion of the off-shell effects incorporated
through spectral functions
(see for example \cite{SCM97,EP1998,KCJKK2002})
can be numerically important.

At large energies one is tempted to completely ignore the effect
of the Fermi motion of nucleons in the colliding nuclei.
According to our knowledge, however, no quantitative estimates have been
presented so far in the literature. 

In the present Letter we wish to show that the Fermi motion effect
on $R_{AA}(p_t)$ is not
negligible even at high energies (W $>$ 15 GeV) and must be included into
consideration, e.g. in the context of the discussion on the creation
of quark-gluon plasma.

\section{Folding model}

In order to demonstrate our main point, in the following we shall
neglect all nuclear effects except the nucleon Fermi motion.
We assume that the collision of nuclei $A_1(Z_1,N_1)$ and
$A_2(Z_2,N_2)$ can be considered as
the collision of two ensambles of nucleons. In general, the production
cross section of a hadron $h$ depends on the type of the collision
($\alpha,\beta$ = pp, pn, np or nn).
We assume that at these high energies all off-shell effects
can be neglected.
In this somewhat simplified picture the cross section {\it per elementary
  collision} 
\footnote{This corresponds to
$ \frac{ \frac{d \sigma^{AA \to h}}{dy d^2P_t} }{N_{bin}}$
in the definition of the nuclear modification factor (\ref{R_AA}).} is
\begin{equation}
\begin{split}
&\frac{d \sigma^{AA \to h}}{d y d^2 p_t}(y,p_t;W) \approx
\int d^3 p_1 d^3 p_2
\sum_{\alpha,\beta=p,n} \frac{N_{\alpha,\beta}}{N_{tot}}
 \; f_{\alpha/1}(p_1) f_{\beta/2}(p_2) \\
&\int d y^* d^2 p_t^* \; \delta(\tilde y-y) \; \delta^2(\vec{\tilde p}_t - \vec{p}_t)
\;
\frac{d \sigma^{\alpha \beta \to h}}{d y^* d^2 p_t^*}(y^*,p_t^*;W^*) \; .
\label{master_formula_1}
\end{split}
\end{equation}
In the above equation $y^*$, $p_t^*$, $W^*$ denote rapidity,
transverse momentum and
energy in the nucleon-nucleon center-of-mass system, respectively.
The quantities $\tilde y$ and $\vec{\tilde p_t}$ (overall CM system)
are obtained
from $\vec{p}_1$, $\vec{p}_2$, $y^*$, $\vec{p}_t^*$, etc., via
appropriate Lorentz transformations and rotations of the frame of reference.
In Eq.(\ref{master_formula_1}) $p_1$ is the
nucleon momentum in the rest frame of the nucleus $A_1$,
while $p_2$ is the nucleon momentum in the rest frame of
the nucleus $A_2$.
Eq.(\ref{master_formula_1}) is valid provided
the standard normalization of momentum distributions is used
\begin{equation}
\int f_{\alpha/i}(p_i) \; d^3 p_i = 1 \; .
\label{normalization_condition}
\end{equation}
In the master formula (\ref{master_formula_1}), $N_{\alpha,\beta}$
is the effective number of the collisions
between nucleons $\alpha$ and $\beta$.
Obviously $N_{tot} = \sum_{\alpha,\beta} N_{\alpha,\beta}$.
In the most naive approach
$N_{\alpha,\beta}/N_{tot} = (N_{\alpha/1} N_{\beta/2}) / ( A_1 A_2 )$.

In our intentionally simplified approach
we consider the nucleus-nucleus collision as a collision
of two bunches of nucleons. For simplicity we make no
distinction between proton-proton, proton-neutron, neutron-proton
and neutron-neutron collisions 
\footnote{This means that our present approach
is valid only for the cases when:
$\sigma_{pp \to h} = \sigma_{pn \to h} =
 \sigma_{np \to h} = \sigma_{nn \to h}$.
This is roughly true for example for h= $\pi^0$, $\pi^+ + \pi^-$,
$\eta$.}.
In addition, we shall make simplifications of kinematics
relevant at high energies, when collective velocities
of both nuclei are much larger than the Fermi velocity, i.e.
$v_{A_1/CM}, v_{A_2/CM} \gg v_F$.
In this case $ \vec{v}_{1/NN},\vec{v}_{2/NN} \; || \; \hat z
\; || \; \vec{v}_{A_1/CM}, \vec{v}_{A_2/CM}$ and $y^* \approx y$.
Then, the invariant distribution (per elementary collision)
of hadron $h$ in nuclear collisions is given as
\begin{equation}
\frac{d \sigma^{AA \to h}}{d y d^2 p_t}(y,p_t;W) \approx
\int d^3 p_1 d^3 p_2 \; f_1(p_1) f_2(p_2) \;
\frac{d \sigma^{NN \to h}}{d y^* d^2 p_t^*}(y^*,p_t^*;W^*) \; .
\label{master_formula_2}
\end{equation}
In the formula above, $W^*$ denotes the energy available
in the elementary nucleon-nucleon collision.
In order to calculate $W^*$ one needs to make the appropriate
Lorentz transformations from four momenta
in the rest frame of nuclei $(E_{i/A_i},\vec{p}_{i/A_i})$
to four-momenta in the overall center of mass of the colliding nuclei
$(E_{i/CM},\vec{p}_{i/CM})$.
Then the nucleon-nucleon c.m. energy is calculated as
\begin{equation}
W^* = \sqrt{ \left( E_{1/CM}+E_{2/CM} \right)^2 
            - \left( \vec{p}_{1/CM} + \vec{p}_{2/CM} \right)^2  } \; .
\label{W_NN}
\end{equation}
In the equation above the on-shell relation
$E_{i/CM} = \sqrt{p_{i/CM}^2 + m_N^2}$ ($i$=1,2) is used.
Due to plateau at midrapidities, at $y \approx$ 0 it seems
reasonable to make an extra approximation $y^* = y$.

Although the master formula (\ref{master_formula_2}) is very simple,
its practical application requires knowledge of nucleon
momentum distributions ($f_1$ and $f_2$) in nuclei as well as
the rapidity, transverse momentum and energy dependence of
the elementary production mechanism.
It is needless to say that realistic calculation of both nucleon
momentum distribution in nuclei and differential cross section for
elementary collisions is rather difficult. While the first requires
nuclear many-body approach to the nucleus structure, the latter involves
nonperturbative QCD effects which are not fully understood.
The exact nuclear calculation based on modern nucleon-nucleon
interactions are available only for the deuteron, $^3H$, $^3He$, and
quite recently -- the four-nucleon system $^4He$ \cite{NKGB2002}.
Similarly, our understanding of the soft sector
(small transverse momenta) of particle production within QCD
is still not complete \cite{szczurek2003}.

\section{Estimate of the effect}

Before we go to the presentation of predictions based on
Eq.(\ref{master_formula_2}), we shall shortly describe
the ingredients which enter the formula.
In the Fermi gas model
\begin{equation}
f(p) = C_{FG} \; \theta(p_F - p) \; ,
\label{Fermi_gas}
\end{equation}
where $p_F$ is Fermi momentum and $\theta$ is the standard
step function. The normalization constant $C_{FG}$
can be obtained from the normalization condition
(\ref{normalization_condition}).
In numerical calculations we take $p_F$ = 1.37 fm$^{-1}$.
In the present paper we shall also consider a simple Gaussian
parametrization of the momentum distribution of nucleons
in $^{12}C$ nucleus from Ref.\cite{CDJKRSS94}.
While the first seems more adequate for heavy nuclei, the latter
may be representative for light nuclei.

The elementary cross sections for the production of $\pi^{\pm}, \pi^0,
K^{\pm}$ in the proton-proton or proton-antiproton collisions are
know experimentally in the broad range of energies. 
As mentioned above, in the present analysis we shall limit ourselves to
the production of $\pi^0$ only. For our purpose a simple parametrization
of the available world data would be useful. Such a parametrization exists,
however, only for charged particles in the limited range of energies
\cite{ISR_parametrization}.
In the parton picture, assuming isospin symmetry, the invariant
cross section for the $ pp \to \pi^0 X$ reaction can be expressed as
\begin{equation}
\frac{d \sigma^{pp \to \pi^0}}{d y d^2 p_t} \left( y,p_t;W \right) =
\frac{1}{2} \left(
\frac{d \sigma^{pp \to \pi^+}}{d y d^2 p_t} \left( y,p_t;W \right) +
\frac{d \sigma^{pp \to \pi^-}}{d y d^2 p_t} \left( y,p_t;W \right)
\right) \; .
\label{sigma_pp_pi0}
\end{equation}
The effects which violate the above relation are expected
to be sufficiently small. In the present calculation we
use a parametrization of the ISR data from
Ref.\cite{ISR_parametrization}.

In general, there are three effects contained in the original convolution
formula (\ref{master_formula_1}) which may result in Fermi motion
modifying of $R_{AA}^h$.
The first effect is connected to the fact that
$p_{t}^* \ne p_{t}$. The second effect is related to the fact that
$y^* \ne y$.
Because the general Lorentz transformation from the NN system to
the overall nucleus-nucleus CM system includes also rotations of
the frame axes, the formulae are rather cumbersome and not easy to reverse.
Therefore these two effects cannot be included exactly in a simple
way. The Monte Carlo method seems more suitable for this purpose.
A Monte Carlo calculation, to be discussed in details elsewhere,
shows that while the first effect is rather small, the second effect
is almost negligible.
 
\begin{figure}[htb] 
  \subfigure[]{\label{fig_W_20_spread}
    \includegraphics[width=7.0cm]{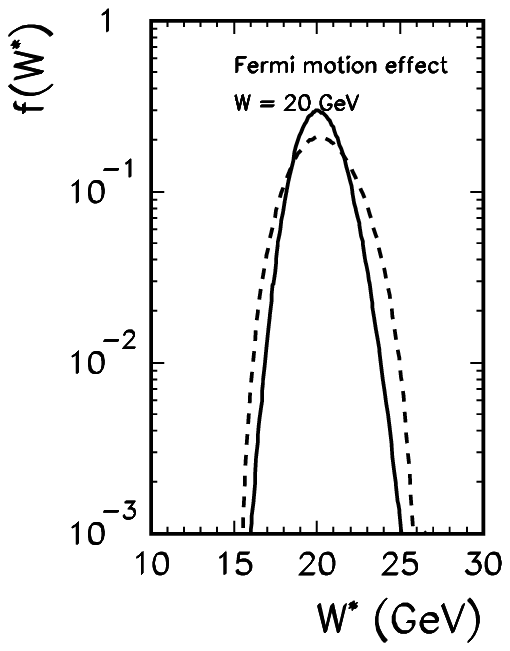}}
  \subfigure[]{\label{fig_W_50_spread}
    \includegraphics[width=7.0cm]{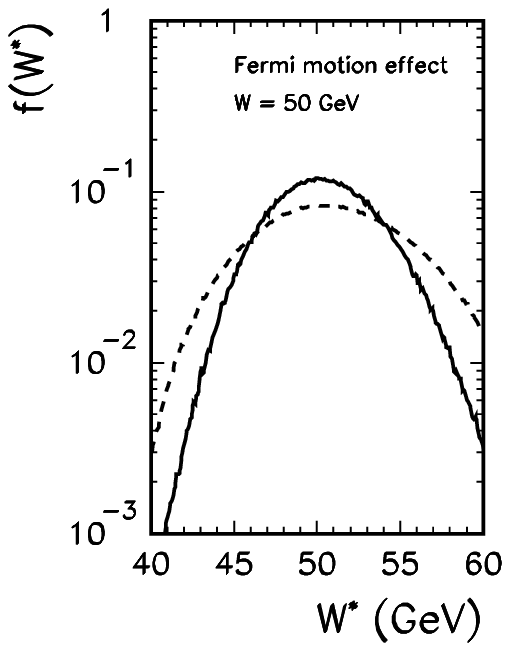}}
\caption{\it
Distribution of the true nucleon-nucleon c.m. energies
for Fermi gas model (dashed) and the Gaussian parametrization
(solid) for nominal energy W = 20 GeV (left panel)
and W = 50 GeV (right panel).
\label{fig:W_spectrum}
}
\end{figure}

\begin{figure}[htb] 
\begin{center}
\includegraphics[width=8cm]{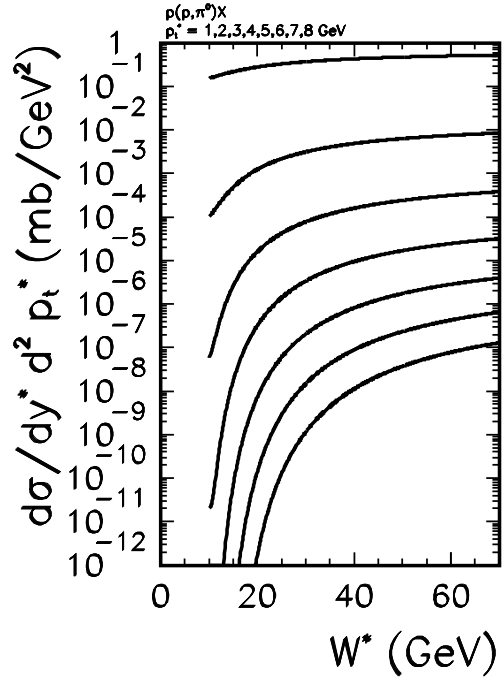}
\caption{\it
The energy dependence of the invariant elementary
cross section for the reaction $ p + p \to \pi^0 + X$ for
different $p_t$ (=1,2,3,4,5,6,7,8 GeV) and $y$ = 0.
\label{fig:pt_W}
}
\end{center}
\end{figure}

The main (third) effect is connected to the fact that
the true c.m. energy in the elementary nucleon-nucleon collision
is not equal to the nominal nucleon-nucleon energy $W^* \ne W$,
even when averaged over Fermi distributions.
As an example, in Fig.\ref{fig:W_spectrum} we show two distributions of
the true nucleon-nucleon center of mass energy $f(W^*)$
\footnote{$\int f(W^*) \; d W^*$ = 1.}
for two
nominal energies $W$ = 20 GeV (left panel) and $W$ = 50 GeV (right panel).
We observe a broadening of $W^*$ distribution with raising nominal energy,
although the relative dispersion $\Delta W^* / W$ stays roughly constant.
The invariant elementary cross sections for particle
production rise across the region over which $f(W^*)$ is spread
(see Fig.\ref{fig:pt_W}) leading to the value of
$\langle W^* \rangle_{\sigma}$
\footnote{The index $\sigma$ indicates
averaging of $W^*$ with the energy-dependent
$NN \to h$ cross section.} larger than $W$.
This effect is stronger for larger $p_t^*$
and smaller energies $W^*$.

\begin{figure}[htb] 
\begin{center}
\includegraphics[width=8cm]{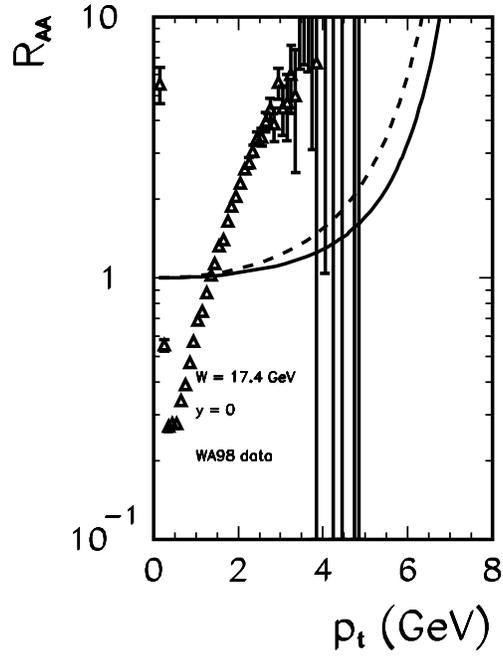}
\caption{\it
The nuclear modification factor $R_{Pb,Pb}^{\pi^0}$
for $W$ = 17.4 GeV and $y$ = 0. The dashed line represent the result for
the Fermi gas momentum distribution, while the solid line
for the Gaussian parametrization of the data for the $^{12}$C nucleus.
The experimental data are from Ref.\cite{WA98_pi0}
for the 12.7 \% most central collisions ($N_{bin}$ = 651).
\label{fig:RAA_WA98}
}
\end{center}
\end{figure}

In Fig.\ref{fig:RAA_WA98} we show $R_{AA}^{\pi^0}$ as a function
of the $\pi^0$ transverse momentum for the $\pi^0$ rapidity y=0
and two different Fermi momentum distributions:\\
(a) Fermi gas model distribution given by
Eq.(\ref{Fermi_gas}) (dashed) and \\
(b) the Gaussian parametrization from Ref.\cite{CDJKRSS94} (solid).\\
The experimental data points were obtained by dividing
the nuclear data from Ref.\cite{WA98_pi0} by the cross section
for the elementary collision obtained with the help of
Eq.(\ref{sigma_pp_pi0}) and the parametrization
from \cite{ISR_parametrization}.

\begin{figure}[htb] 
  \subfigure[]{\label{fig_W_fermi_gas}
    \includegraphics[width=7.0cm]{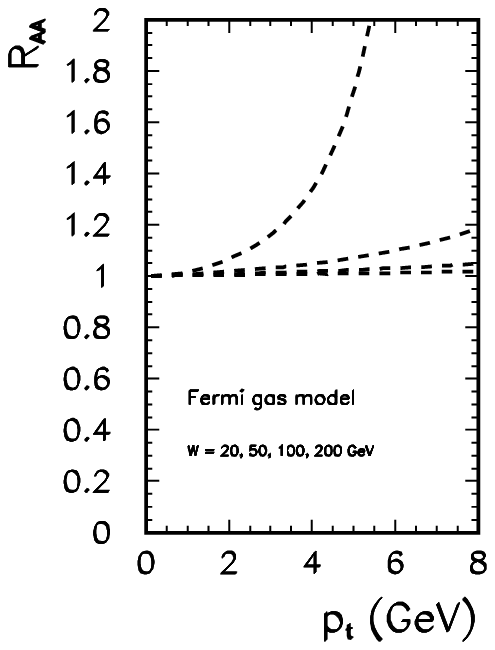}}
  \subfigure[]{\label{fig_W_parametrization}
    \includegraphics[width=7.0cm]{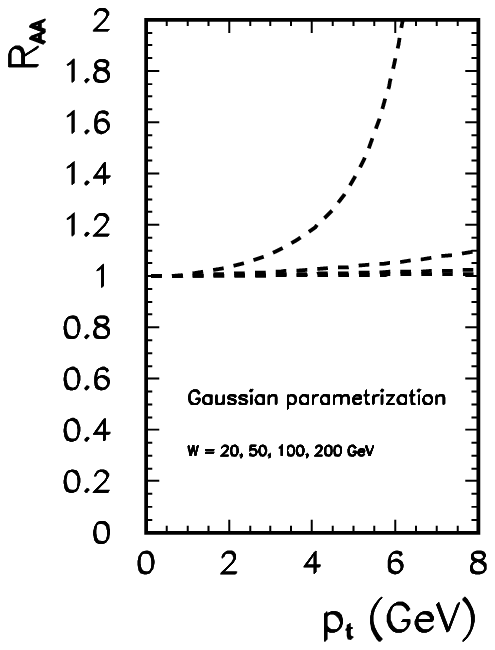}}
\caption{\it
Energy dependence of $R_{AA}^{\pi^0}$ at $y$ = 0.
The lines correspond to the nominal energies $W_{NN}$ =
20, 50, 100, 200 GeV in the Fermi gas model (left panel) 
and using a simple Gaussian parametrization (right panel).
The small fluctuations are due to poor numerics.
\label{fig:RAA_W}
}
\end{figure}

The Fermi motion effect calculated from our master formula
(\ref{master_formula_2}) is strongly energy dependent,
which is demonstrated in Fig.\ref{fig:RAA_W}.
Generally, the larger energy the smaller the effect.
Surprisingly, the effect of the Fermi motion 
stays present even up to W = 50 GeV.

In the present analysis we have estimated the effect of Fermi motion
assuming that nucleons can be treated as free on-mass-shell particles.
Under this assumption the light-cone fraction carried by a nucleon,
$\xi$, defined so that 0$ < \xi < $A, is
$\xi=(\sqrt{m_N^2+p^2}+p_3)/(m_A/A)$. As a result the average
$<\xi>$ is larger than 1 and the momentum sum rule is violated -
nucleons carry momenta larger than the total nucleus momentum.
Effectively the energy in NN collisions is overestimated.
This is connected with the effect of nucleon binding in nuclei.

The problem can be even better seen in the nucleus rest frame,
where the total nucleon-gas energy is larger than the nucleus mass
\begin{equation}
A \cdot \int \sqrt{p^2+m_N^2} \; f(p) \; d^3 p \; > \; M_A \; ,
\label{too_much_energy}
\end{equation}
i.e. the gas of nucleons has too large energy. This effect can be 
cured introducing effective nucleon mass $m_N^{eff}$:
\begin{equation}
A \cdot \int \sqrt{p^2+(m_N^{eff})^2} \; f(p) \; d^3 p \; = \; M_A \; . 
\end{equation}
This corresponds to the reduction of the nucleon mass by 10-20\%.

\begin{figure}[htb] 
\begin{center}
\includegraphics[width=10cm]{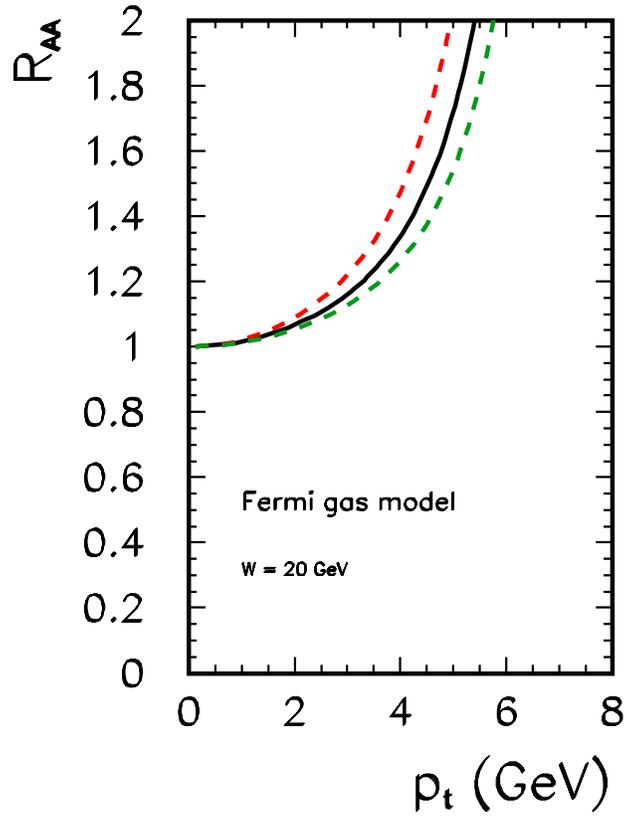}
\caption{\it
The effect of the effective nucleon mass variation for W=20 GeV.
The solid mass is for standard free nucleon mass while the dashed lines
are for nucleon mass modified by $\pm$ 150 MeV.
In this calculation $y$ = 0.
\label{fig:binding}
}
\end{center}
\end{figure}

Can the binding effect change our conclusions?
The binding effect in nuclei is in general a complicated multiparticle
problem and was solved in the exact way essentially only for
three- and four-nucleon systems.
The binding effect in heavy nuclei can be approximately
included by introducing a notion of nucleon effective mass
in the relation $E_{N/A} = \sqrt{p_{N/A}^2 + (m_N^{eff})^2}$.
Since the binding effect is only a correction to our main effect
of Fermi motion this approach seems justified.
For illustration in Fig.\ref{fig:binding} we show the result
for nominal energy W = 20 GeV
with nucleon mass modified by $\pm$ 150 MeV. This figure shows
that the binding effect is not only qualitatively but also
quantitatively a correction to our effect.
This correction does not change the main trend of
$R_{AA}(p_t)$. A more refined calculation is not our intention and
clearly goes beyond the scope of the present paper.

\begin{figure}[htb] 
\begin{center}
\includegraphics[width=10cm]{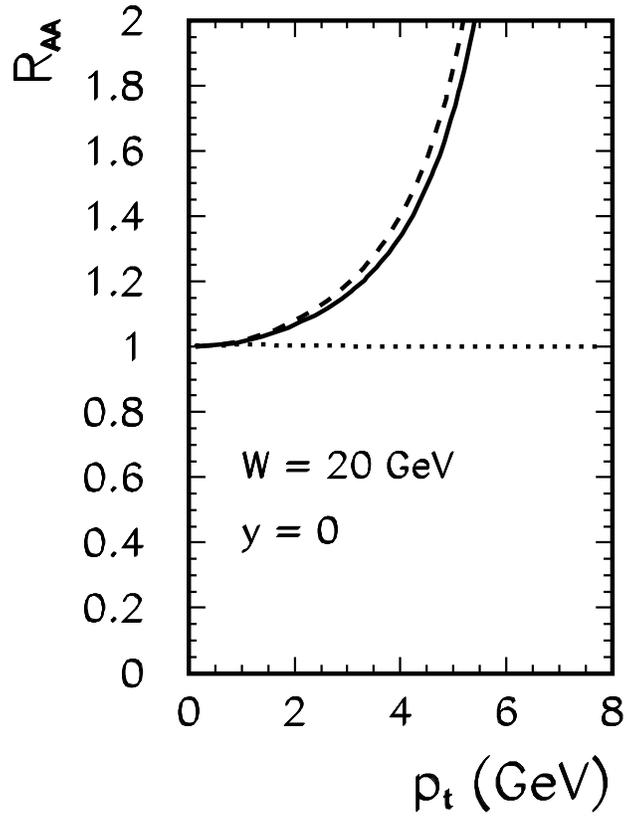}
\caption{\it
Fermi motion effect on nuclear modification factor for
$\pi^+ + \pi^-$ (solid), $K^+ + K^-$ (dashed) and $p + \bar p$ (dotted)
at $W$ = 20 GeV and $y$ = 0. In this calculation Fermi gas model was used.
\label{fig:RAA_particles}
}
\end{center}
\end{figure}

How does the effect of Fermi motion depend on the type of particle
measured? In Fig.\ref{fig:RAA_particles} we present $R_{AA}$
for $\pi^+ + \pi^-$ (solid), $K^+ + K^-$ (dashed) and
$p + \bar p$ (dotted) at the center-of-mass energy $W$ = 20 GeV
and $y$ = 0.
In this calculation Fermi gas model was used.
The results for pions and kaons are similar, while the result
for protons+antiprotons are somewhat different.
The latter does not show the enhancement at large transverse
momenta observed for pions and kaons. This is due to the
parametrization for protons and antiprotons \cite{ISR_parametrization}
which is practically energy independent.
Perhaps the difference for protons+antiprotons and pions/kaons
can be used to identify experimentally the Fermi motion effect.

In the present letter we have concentrated only on the effect
of Fermi motion on nuclear modification factor.
The Fermi motion in colliding nuclei is probably
important in many other fixed-target (low energy) processes.
For example global partonic QCD analyses use Drell-Yan data
based on rather low-energy proton-nucleus collisions
than elementary ones.
The standard QCD approach is known to have problems in describing
the transverse momentum distribution of the lepton pair.
In such analyses the Fermi motion of nucleons
in nuclei is completely ignored.
We expect that the effect of the Fermi motion should be
dependent on the invariant mass of the two leptons.
In general, the larger invariant mass, the larger
Fermi motion effect may be expected.
From the experience of the present paper one may
expect large effects close to kinematical limits.
Therefore even for the production of heavy gauge bosons (W,Z)
in nucleus-nucleus collisions at RHIC one may expect
sizeable effect due to the Fermi motion.
These effects were never discussed in the literature
and await a quantitative analysis.

Finally we wish to mention another interesting, not fully
explained, effect related to the CERES collaboration measurement
of the elliptic flow parameter $v_2$ \cite{CERES}.
The existing hydrodynamical or thermodynamical models have
serious problems in explaining a saturation of $v_2$ observed
at larger transverse momenta (see e.g.\cite{NATO}).
It is quite probably that the Fermi motion effect may play
important role in the full understanding of this effect.
However, this effect is very subtle and requires the inclusion
of many effects in addition to the nucleon Fermi motion.
This clearly goes beyond the scope of the present, intentionally
simplified, analysis. 

\section{Conclusions}

For the first time in the literature, in the present Letter
we have analysed the role of the Fermi motion in nuclei
in modyfing the so-called nuclear modification factor $R_{AA}$.
A large effect is found even at SPS energies.
The details depend on the model of the Fermi motion as well
as on the cross sections for elementary collisions. The latter
are very difficult to calculate reliably, especially at
low transverse momenta. Therefore we have made use of
a parametrization of existing experimental data.

At SPS energies the Fermi motion causes a sizeable enhancement at
$ p_t >$ 4 GeV. At RHIC energies the enhancement almost dissapears.

This quantitative difference of the role of the Fermi
motion at low and high energies is very important in
the context of a quantitative understanding of jet quenching
in dense nuclear media like quark-gluon plasma.
At present the latter cannot be calculated from first principles
and the model parameters are extracted by comparison
to the experimental data.

Even when alone, the Fermi motion leads to the Cronin effect.
Different mechanisms like parton multiple scattering
\cite{parton_rescattering}, initial transverse momentum broadening
in nuclei \cite{Wang2000,ZFPBL2002} or perturbative saturation
of gluon distributions \cite{BKW03} 
can contribute to the Cronin effect too.
Our analysis shows that the Fermi motion effect is very important
and must be included into the analysis before reliable conclusions
about other nuclear effects may be drawn from such phenomenological studies.
We find $R_{AA}^{\pi^0}(RHIC) < R_{AA}^{\pi^0}(SPS)$ which makes
the interpretation of the experimental data even more complicated.
For example at the SPS energy W=17.4 GeV the inclusion of the Fermi motion
would improve the agreement of the calculation in Ref.\cite{VG02}
with the WA98 collaboration experimental data (see their Fig.3).
However, the inclusion of all effects like shadowing, quenching
and Fermi motion in one consistent framework seems
rather difficult at present.

The Fermi motion effect seems to be one of a few different
competing mechanisms modyfing $R_{AA}$.
Can one isolate experimentally the Fermi motion effect better?
We think that the role of Fermi motion can be pinned down
better in the collisions of light nuclei, where other effects are suppressed
and the nucleon momentum distribution can be calculated from
realistic models of nucleon-nucleon interactions.
The examples are $d - d$ or $\alpha-\alpha$ scattering measured
already in the past (see for example \cite{BCMOR_RAA}).
Better precision data at SPS energies would
help to identify the effect of the Fermi motion via a comparison
with realistic theoretical calculations \cite{NKGB2002}.
Also very precise data on elementary collisions would be
helpful when calculating the modification of $R_{AA}$ precisely.

\vskip 1cm

{\bf Acknowledgments}
We are indebted to Klaus Reygers for providing us with
the WA98 collaboration experimental data for
$\pi^0$ production in $^{208}Pb + ^{208}Pb$ collisions
and Zbigniew Rudy for providing us with a parametrization
for the nucleon momentum distribution in $^{12}C$.
One of us (A.S.) is indebted to Piotr Bo\.zek and Wojciech Florkowski
for an interesting discussion and Piotr \.Zenczykowski for
his careful reading the manuscript.
We are indebted to Kolya Nikolaev and Andrzej Rybicki for
a discussion which helped in finding a mistake in our code
and therefore helped in correcting a first version of our letter.



\newpage


\end{document}